\documentclass[a4paper]{jpconf}
\usepackage{graphicx}
\usepackage{amssymb}
\begin{document}
\title{A Determination of the Luminosity Function of DA White Dwarfs
from the Kiso Survey}

\author{M-M Limoges and P Bergeron}

\address{D\'epartement de Physique, Universit\'e de Montr\'eal, C.P.~6128, 
Succ.~Centre-Ville, 
Montr\'eal, Qu\'ebec H3C 3J7, Canada}

\ead{limoges@astro.umontreal.ca, bergeron@astro.umontreal.ca}

\begin{abstract}
We rederive the luminosity function for the sample of DA white dwarfs
from the Kiso Schmidt ultraviolet excess survey (KUV stars) using the
spectroscopic method where the atmospheric parameters ($T_{\rm eff}$
and $\log g$) and absolute visual magnitudes for each star are
obtained from detailed model atmosphere fits to optical spectroscopic
data. We compare the result of our determination with that obtained by
Darling (1994) based on empirical photometric calibrations. Our
luminosity function is also compared with that derived
spectroscopically from the PG survey. Misclassified objects are also
briefly discussed.\end{abstract}

\section{Introduction}
The Palomar Green (PG) survey has identified 1874 ultraviolet excess
objects and it is estimated to be 84\% complete (Green et
al.~1986). The luminosity function of DA white dwarfs from the PG
survey was determined from photometric calibrations by Fleming et
al.~(1986) and improved by Liebert et al.~(2005; hereafter LBH05)
using the spectroscopic method for measuring effective temperatures,
surface gravitites, and absolute visual magnitudes (see, e.g.,
Bergeron et al.~1992). Similarly, Darling (1994) derived a luminosity
function of all white dwarfs from the Kiso Schmidt ultraviolet excess
survey (KUV) --- all spectral types included --- using $M_V$ values
obtained from photometric calibrations. However, the $B-V$ versus
(photographic) color index relation used to derive the absolute visual
magnitudes showed a considerable dispersion, and it is believed that
the estimated $M_V$ values probably reflect a similar dispersion
(LBH05). LBH05 (see their Fig.~10) made a detailed comparison of their
luminosity function with those of Fleming et al.~(1986) and Darling
(1994). They evaluate the completeness of the PG survey to 75\%, while
Darling (1994) found a lower value of 58\%. In this paper, we present
preliminary results from our study aimed at improving the luminosity
function by applying the spectroscopic method to the DA
stars in the KUV sample. One of our goals is to improve the
comparison of the DA luminosity functions derived from the PG and KUV
samples.

\section{White Dwarfs from the KUV Survey}
The Kiso Ultraviolet Excess Survey (KUV) is a photometric search for
UV-excess objects, performed with the 105-cm Schmidt telescope at Kiso
Observatory. The 1186 objects were found in 44 fields in a belt from
the northern to the southern galactic pole at a galactic longitude of
$180\,^{\circ}$. The fields of the survey have a mean limiting
magnitude of 17.7 and cover a total area of 1400 square degrees
(Noguchi et al.~1980, Kondo et al.~1984).

We have secured optical spectroscopic observations for all white
dwarfs in the KUV survey. Spectra have been obtained with the
Steward Observatory 2.3-m telescope equipped with the Boller
\& Chivens spectrograph and a Loral CCD detector. The 4.5 arcsec slit
together with the 600 l/mm grating in first order provided a spectral
coverage of $\lambda\lambda$3200--5300 at an intermediate resolution
of $\sim 6$~\AA\ FWHM and a signal-to-noise ratio of at least 50. We
have reclassified several objects (see Table 1): 30 objects from
Darling's sample turned out to be main sequence or subdwarf stars. Our
analysis also includes seven new white dwarfs that have been
identified in the KUV survey since 1994. Only DA stars are analyzed
here, for a total of 172 white dwarfs (including the 5 magnetic
DAs). Sample spectra are displayed in Figure 1.

\begin{table}[h]
\caption{Spectroscopic content of the KUV survey.}
\begin{center}
\begin{tabular}{l|cccccc}
\br
  &DA&DB&DQ/DZ/DC&DO&Mag DA&Total\\
\mr
This work&167&23&15&1&5&211\\
Darling&192&25&15&0&2&234\\
\br
\end{tabular}
\end{center}
\end{table}

\begin{figure}[h]
\begin{center}
\includegraphics[width=30pc]{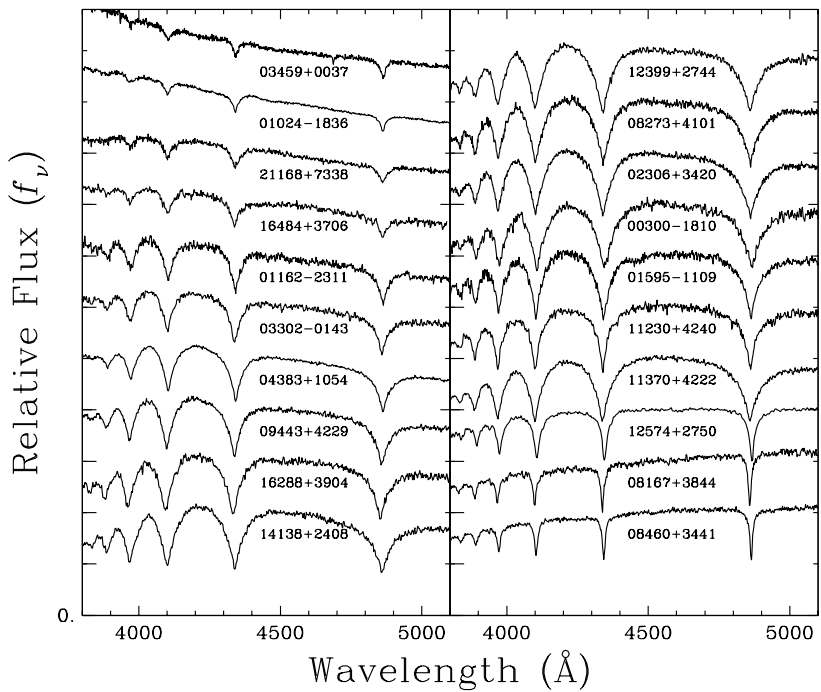}
\end{center}
\caption{Optical spectra for a subset of DA stars from the KUV 
sample. The effective temperature decreases from upper left to bottom right.}
\end{figure}

With our sample properly defined, we proceed to measure the
atmospheric parameters ($T_{\rm eff}$ and $\log g$) of each star using
the spectroscopic fitting technique and model atmospheres described at
length in LBH05 and references therein. Our results for the KUV sample
are summarized in Figure 2 in a $\log g$ vs $\log T_{\rm eff}$ diagram.
Absolute visual magnitudes are then calculated from evolutionary
models following the prescription of Holberg and Bergeron (2006). Our
spectroscopic $M_V$ values are compared in Figure 3 with those derived
by Darling (1994), which are based on empirical photometric
calibrations. The differences observed are quite significant, with the
empirical estimates of Darling being generally fainter than the
$M_V$ values determined spectroscopically. It is therefore expected
that the luminosity function based on these absolute magnitudes will
be affected as well.

\begin{figure}[h]
\begin{center}
\includegraphics[width=30pc]{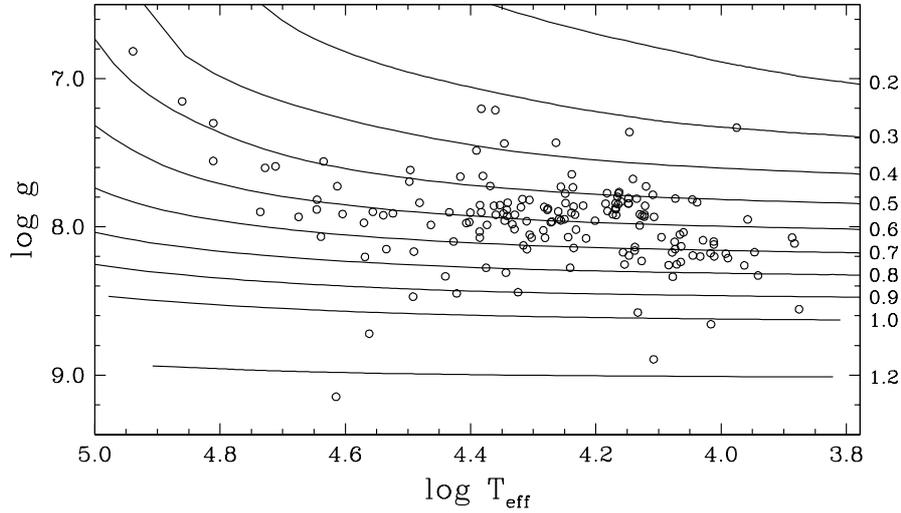}
\end{center}
\caption{$T_{\rm eff}$ and $\log g$ values for all DA stars 
from the complete KUV sample. The solid lines represent carbon-core
evolutionary models with thick hydrogen layers (see LBH05 and references therein);
numbers on the right hand side of the figure indicate the mass of each
model in solar masses.}
\end{figure}

\begin{figure}[h]
\begin{minipage}{18pc}
\hspace{1.3pc}
\includegraphics[width=20pc]{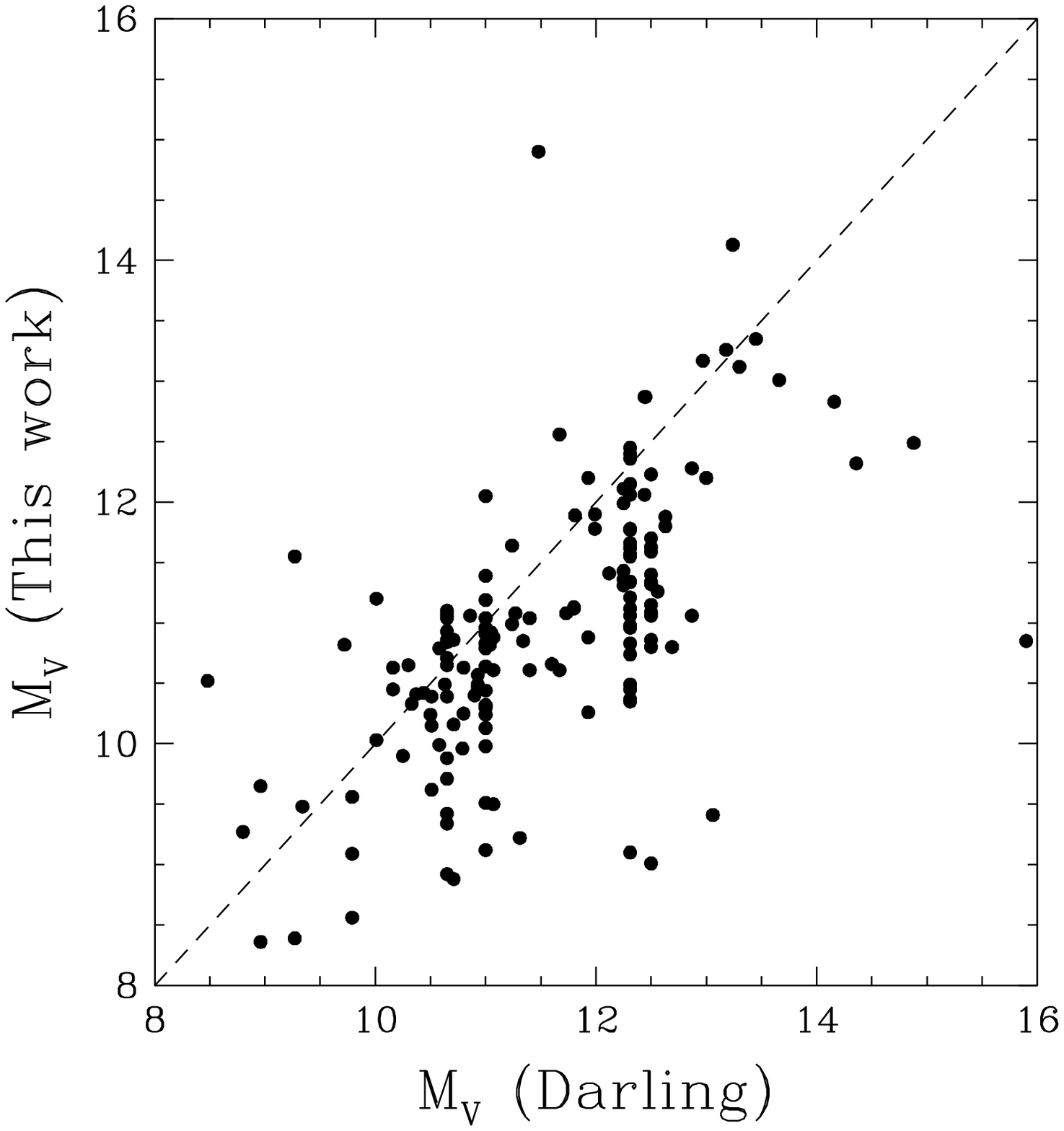}
\caption{Comparison of the absolute visual magnitudes 
obtained from the empirical photometric calibration 
of Darling (1994) and from the spectroscopic method.}
\end{minipage}\hspace{2pc}%
\begin{minipage}{18pc}
\hspace{1pc}
\includegraphics[width=20pc]{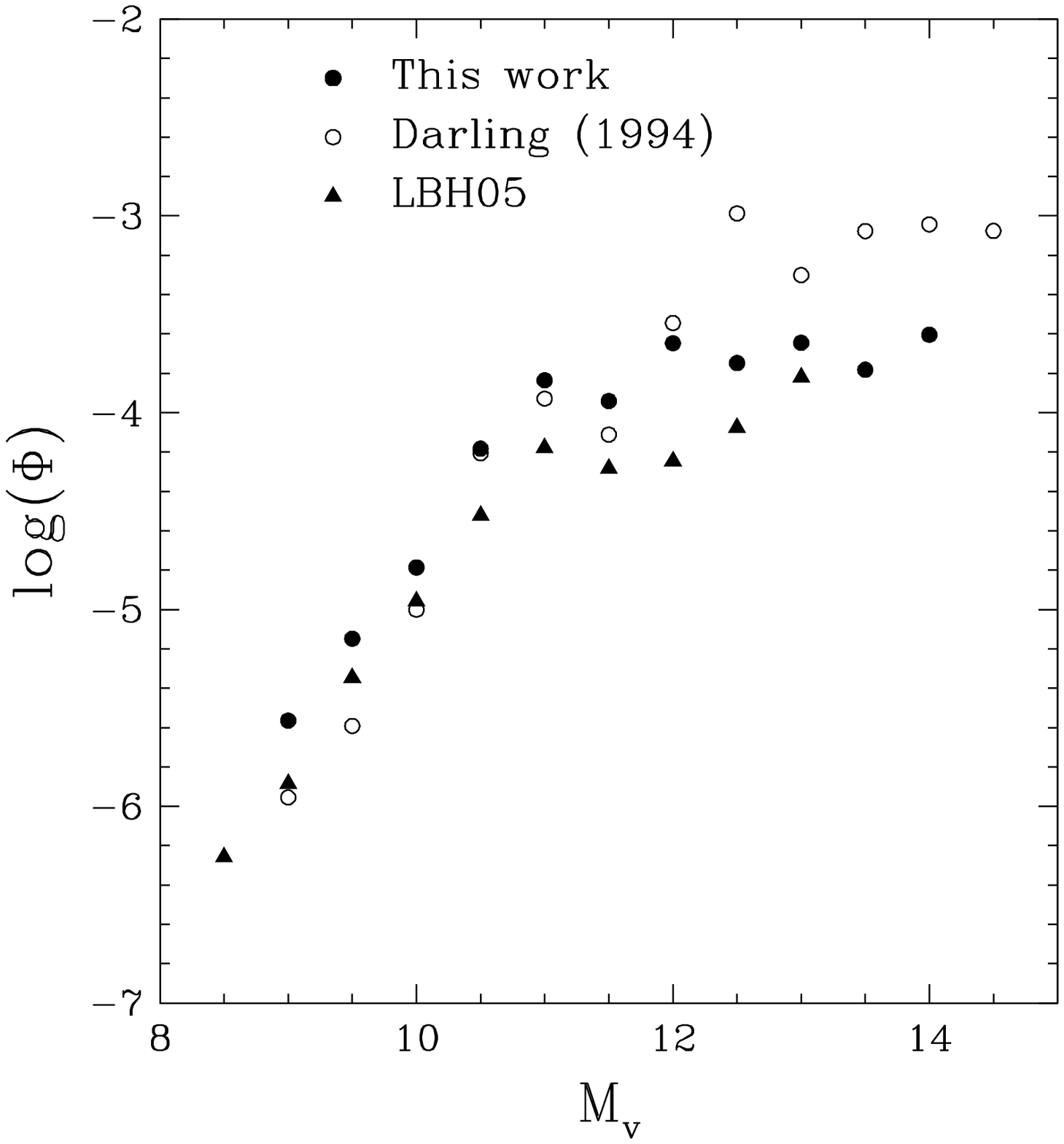}
\caption{Comparison of the luminosity functions of DA 
stars from the KUV survey (this work and Darling 1994) and 
from the PG survey (LBH05).}
\end{minipage} 
\end{figure}

\section{The Luminosity Function}
The luminosity function of white dwarf stars is a measure of the local
density of this type of object by interval of absolute magnitude. It
contains information about the history of the local galactic disk and
stellar evolution. It represents also a direct measure of the space
density of white dwarfs and of the stellar death rate in the local
galactic disk. Compared to theoretical luminosity functions, it can
lead to an estimate of the age of the local galactic disk, as
discussed in Fontaine et al.~(2001) for instance. We derive the
luminosity function for the DA stars from the KUV survey by adding the
inverse of the contribution of each star to the space density for each
magnitude bin, following the well established $1/V_{\rm max}$ method
(Schmidt 1968). The completeness of the sample is evaluated from the
$\left< V/V_{\rm max}\right>$ method described in Green (1980). Using
this technique, we obtain a limiting magnitude of 17.31 for our sample
of DA stars.

Our improved luminosity function for the KUV sample based on
spectroscopic determinations of $M_V$ is compared in Figure 4 (filled
circles) with the results of Darling (1994) based on empirical
photometric calibrations (open circles). Also shown is the luminosity
function for the DA stars from the PG survey derived by LBH05 using
the same spectroscopic technique as that used in our analysis (filled
triangles). Our results for the KUV sample are systematically larger
than the PG sample in each magnitude bin, which suggests that the KUV
survey is more complete than the PG survey.  This is particularly true
at the faint end of the luminosity function where the PG sample is
known to be fairly incomplete (LBH05). We note, however, that the
differences do not appear as large as previously estimated from the
comparison with the results of Darling (see also LBH05,
Fig.~10). Indeed, above $M_V\sim 12$, our spectroscopic determination
of the luminosity function is systematically lower than that of
Darling. This is explained by the fact that a lot of the objects in
the fainter bins in Darling's analysis have been shifted to brighter
magnitude bins as a result of his overestimates of the $M_V$ values
(see Fig.~3). We must also mention that because of our spectral
reclassification (see above), our sample contains fewer DA white
dwarfs than the original sample analyzed by Darling. According to
Darling, the completeness of the PG survey is 58\%, while the KUV
survey is 75\% complete. The total space density of DA stars based on
our analysis of the KUV sample is $2.10\times10^{-3}$ pc$^{-3}$. LBH05
obtained for the PG survey a value of $0.5\times10^{-4}$ pc$^{-3}$ for
DA white dwarfs with $M_V< 12.75$. For the same range of
$M_V$, we derive a value of $7.58\times10^{-4}$ pc$^{-3}$, i.e.~a
space density $\sim 15$ times larger. The next step in our analysis
will be to include the DB stars observed in our survey to improve our
determination of the complete luminosity function of white dwarf
stars.

\ack
We would like to thank the director and staff of Steward Observatory
for the use of their facilities. This work was supported in part by
the NSERC Canada and by the Fund FQRNT (Qu\'ebec). P.B. is a Cottrell
Scholar of the Research Corporation for Science Advancement.

\section{References}
\medskip
\begin{thereferences}
\item Bergeron, P, Saffer, R and Liebert, J 1992, {\it ApJ} {\bf 394} 247
\item Green, R F 1980 {\it ApJ} {\bf 238} 685 
\item Green, R F, Schmidt, M and Liebert, J 1986, {\it ApJS} {\bf 61} 305
\item Darling, G W 1994 {\it PhD Thesis}, Dartmouth College, Hanover NH
\item Fleming, T A, Liebert, J and Green, R F 1986, {\it ApJ} {\bf 308} 176
\item Fontaine, G, Brassard, P and Bergeron, P 2001 {\it PASP} {\bf 113} 409 
\item Holberg, J B and Bergeron P 2006, {\it ApJ} {\bf 132} 1221
\item Kondo, M, Noguchi, T and Maehara, H 1984 {\it Ann. Tokyo Astron. Obs.} {\bf 20} 130 
\item Liebert, J, Bergeron, P and Holberg, J B 2005 {\it ApJ} {\bf 156} 47 (LBH05)
\item Noguchi, T, Maehara, H and Kondo, M 1980 {\it Ann. Tokyo Astron. Obs.} {\bf 18} 55 
\item Schmidt, M 1968 {\it ApJ} {\bf 151} 393 
\end{thereferences}

\end{document}